\documentclass[journal]{article}
\usepackage[english]{babel}

\usepackage[letterpaper,top=2cm,bottom=2cm,left=3cm,right=3cm,marginparwidth=1.75cm]{geometry}
\usepackage{amsmath}
\usepackage{graphicx}
\usepackage[colorlinks=true, allcolors=blue, breaklinks=true]{hyperref}
\usepackage{xcolor}

\usepackage[
labelfont=sf,
hypcap=false,
format=hang,
]{caption}
\usepackage{capt-of}
\usepackage{amsmath,amsfonts}
\usepackage{algorithmic}
\usepackage{array}
\usepackage[caption=false,font=normalsize,labelfont=sf,textfont=sf]{subfig}
\usepackage{textcomp}
\usepackage{stfloats}
\usepackage{url}
\usepackage{verbatim}
\usepackage{graphicx}
\def\BibTeX{{\rm B\kern-.05em{\sc i\kern-.025em b}\kern-.08em
    T\kern-.1667em\lower.7ex\hbox{E}\kern-.125emX}}
\usepackage{balance}
\usepackage{wrapfig}
\usepackage[style=ieee]{biblatex}
\graphicspath{{../Images}}
\usepackage{xcolor}
\usepackage[pagewise]{lineno}

\title{Analytical approximations for magnetic coupling coefficients between adjacent coils}
\author{Robert R. Hughes\thanks{School of Electronic, Electrical and Mechanical Engineering, Faculty of Engineering, University of Bristol, United Kingdom, BS8 1TR (e-mail: robert.hughes@bristol.ac.uk)}, Alexis {Hernandez Arroyo}*, Anthony J. Mulholland \thanks{School of Engineering Mathematics and Technology, Faculty of Engineering, University of Bristol, United Kingdom, BS8 1TR}
}
\begin{document}
\maketitle

% \linenumbers
\begin{abstract}
This paper presents a simple yet novel two-dimensional modelling approach for approximating the coupling coefficient between neighbouring inductors as a function of co-planar separation and relative angular displacement. The approach employs simple geometric arguments to predict the effective magnetic flux between inductors.  Two extreme coil geometry regimes are considered; planar coils (i.e. on printed circuit board), and solenoid coils, each with asymmetric ferrite cores about the central magnetic plane of the inductor.
The proposed geometric approximation is used to predict the coupling coefficient between sensors as a function of separation distance and angular displacement and the results are validated against two-dimensional finite element modelling results.  The analytical approximations show excellent agreement with the FE analysis, predicting comparable trends with changing separation and angular displacement, enabling best fitting to 2D FE and 3D numerical data with a residual standard deviation of less than $0.5\%$ for the planar coil approximation.
The work demonstrates the validity of the analytical approximation for predicting coupling behaviour between neighbouring coils. This has practical uses for the automated estimation of the physical separation between coils, or the curvature of surfaces they are rested or adhered to.
\end{abstract}

\section{Introduction}

Inductors are found in a diverse range of applications from non-destructive testing (NDT) \cite{Garcia-Martin2011Non-destructiveTesting} to wireless power-transfer (WPT) \cite{Zhong2020BasicWPT}. 
In many of these applications, configurations of multiple coils are used, and the computation of the expected coupling coefficients between coils is of significant interest \cite{Babic2008CalculatingAir, Babic2010MutualFormula, Akyel2009MutualAxes}.
The coupling coefficient is a calculated empirical measure of the amount of magnetic flux sharing between coils and is therefore a parameter that engineers seek to maximise to promote the greatest efficiencies of their systems.  
There are other applications such as in meta-material and microwave antenna design where a clear understanding of the relationship between the angle or separation of adjacent coils, and the coupling factor is desired \cite{Tyurnev2010CouplingTheory}. 
However, the computation of realistic coupling coefficients for arbitrary coil geometries, relative proximities and orientations to one another is non-trivial. 

Solutions to these problems typically rely on either finite element modelling techniques \cite{su2009mutual, acero2011analysis} or the numerical integration of elliptical integrals \cite{Conway2007InductanceFunctions, hussain2022simplified, ravaud2010cylindrical}. There are therefore no closed form analytical solutions for computing relative physical variables as a function of the coupling coefficient.  While there are multiple methods for experimentally calculating the coupling coefficient between coils \cite{Wang2021OptimizationCoils} \cite{Tyurnev2010CouplingTheory}, a direct inversion of physical parameters (i.e. separation and relative angle) from a calculated coupling coefficient is non-trivial.
Many works calculate the coupling coefficient between inductors using circuit theory \cite{Seo2019ComparativeEfficiency}. In this paper the development of a new formula simplified for the simulation of 2D coils is presented using a simplified magnetic flux model.

In this paper, a simplified two-dimensional (2D) approximation is devised for computing the magnetic flux shared between neighbouring coils, employing trigonometric arguments. 
First order formulae are derived to predict the coupling coefficient as a function of separation and relative angle between the coils.  Two distinct coil designs are considered; planar-style coils (i.e. printed circuit board windings), and solenoid-style coils.  The resulting approximations are validated against 2D and 3D finite element models.

\section{Modelling Coupling Coefficients}\label{sec:kmodel}
Determining coupling coefficients between neighbouring coils is an active area of research for many, particularly for those interested in wireless power transfer applications.  Solutions to these problems are non-trivial and often require numerical integration of specific elliptical functions to arrive at analytical formulae.  Details of these can be found in resources such as \cite{Babic2008CalculatingAir}. 
However, in this paper, we present a simplified first-order approximation for predicting the general trends exhibited in coupled two-coil systems. These simple formulae can then be easily employed to fit to calibration data and used to invert the physical displacement (separation of angle) between coils.

\begin{figure*}[!t]
\centering
\includegraphics[width=5.0in]{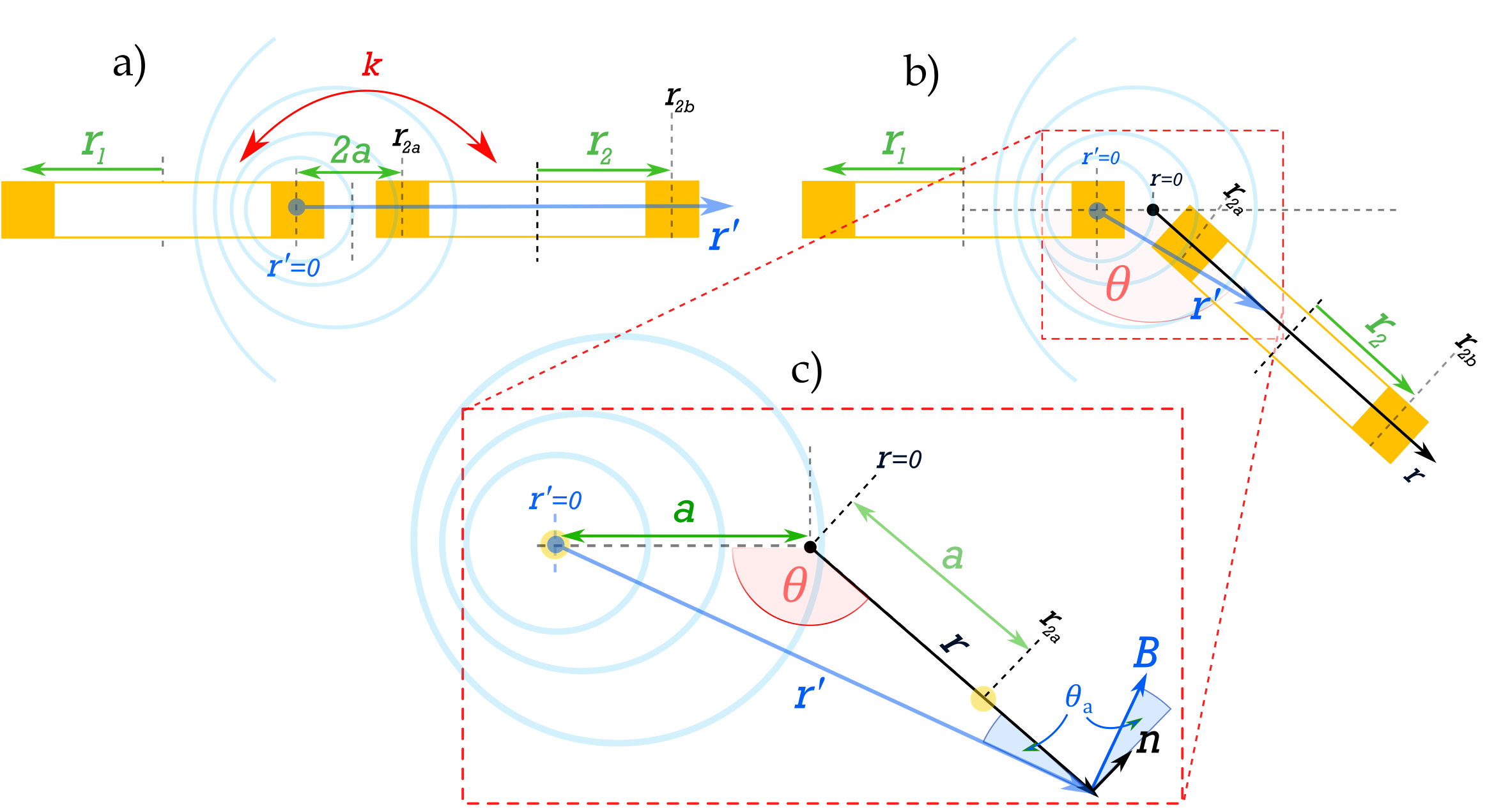}
\caption{Example configuration of two planar filament coils, where the yellow blocks represent the cross-section of the coils of radius $r_n$, with separation $2a$ between the magnetic point source of coil 1 and the nearest winding centre located at a point $r_{2a}$ and coupling $k$. The centre of the furthest winding of coil 2 is at point $r_{2b}$.  The diagrams show; a) a co-planar configuration, b) an angled configuration where $\theta$ is the relative angle between the 2 coil planes, and $\theta_a$ is the angle between normal axis of coil 2 and the B-field. Diagram c) shows the zoomed-in detail of the trigonometric relationship between distance from the pivot point ($r=0$) and the distance from B-field source ($r'=0$).}
\label{fig:coil-geom}
\end{figure*}

To calculate the changes in the coupling coefficient, $k$, due to the geometric configuration of a sensor system, we can consider a simple theoretical model for the magnetic flux between two neighbouring sensors.  Figure~\ref{fig:coil-geom} shows the physical configuration of two coupled resonant coils of radius $r_1$ and $r_2$ respectively. Assuming coil 1 is excited with a current $I_1$, while coil 2 is passive, coil 1 will generate a magnetic flux, $\Phi_1$, that is proportional to the current as, 
\begin{equation}
\Phi_1 \approx \mu n_1 I_1 a_1, \label{eqn:Flux1}
\end{equation}
where $n_1$ is the turn density of the coil, $\mu$ is the permeability of the core and $a_1$ is the planar cross-sectional area of its core. We can express the magnetic flux through coil 2 as \cite{Griffiths2008IntroductionElectrodynamics},
\begin{align}
\Phi_2  &= \int_{a_2} B \cdot \,da, \label{eqn:k_flux0}
%&= \frac{1}{\phi_1} \oint_{s_2} A \cdot \,ds, \label{eqn:k_flux1} \\
% &= \frac{1}{4\pi a_1} \oint \left( \oint \frac{ds_1}{\tau} \right) \cdot \,ds_2. \label{eqn:k_flux2}
\end{align}
where $B$ is the magnetic flux density passing through coil 2, $a_2$ is the cross-sectional area of coil 2, and $da$ is the incremental area.  We can then define simplified expressions for the flux through the secondary coil for semi-infinite coil configurations shown in Figure~\ref{fig:coil-geom}. 

The magnetic flux through coil 2, $\Phi_2$, is the integral of the magnetic flux density generated by coil 1, $B$, over the area enclosed by coil 2, $a_2$.  
% Assuming for now that both coils are infinitely long and identical in length, $l$ (into the page),
Let us consider that coils 1 and 2 have identical filament turns and semi-infinite lengths into the page - equivalent to long narrow coils (i.e. where the radius $r_1 = r_2 = r_c \ll l_c$), where $l_c$ is the length of the coils. In this scenario, the magnetic flux density can be assumed to be the same at all points along the length of the coil. Therefore the coupling coefficient, $k$, can equally be expressed as the ratio between the flux per-unit-lengths ($\Phi_n' = \Phi_n/l_c$) as,
\begin{align}
k \approx \frac{\Phi_2'}{\Phi_1'} &= \frac{1}{\Phi_1'} \int_{s_2} B \cdot \,ds, \label{eqn:k_flux1}
\end{align}
where $ds$ is the incremental length across the plane of coil 2, and $s_2$ is equivalent to the width of coil 2. Assuming, for an infinitely long coil, the magnetic flux density decays as $1/(2\pi r)$ away from the current source (as for around a line current), and that the total flux to one side of the excitation coil must be equal to half the total flux inside the coil \cite{Griffiths2008IntroductionElectrodynamics}, the magnetic flux density to the side of coil 1 is approximated to,
\begin{equation}
B(r') \approx \frac{\Phi_1'}{4\pi r'}, \label{eqn:Br}
\end{equation}
where $r'$ is the radial distance from the magnetic field source (centre of magnetism) within the windings. Equation~\ref{eqn:k_flux1} therefore gives,
\begin{align}
       k \approx \frac{1}{4\pi} \int_{r'_{2a}}^{r'_{2b}} \frac{1}{r'} \,dr' = \frac{1}{4\pi} \ln{\left[\frac{r'_{2b}}{r'_{2a}}  \right]}, \label{eqn:k_gen}
\end{align}
where $r'_{2a}$ and $r'_{2b}$ are the radial distances from the magnetic point source (at $r'=0$) to the nearest and furthest windings of coil 2 respectively, and can be defined for any separation $a$, coil height, $h$, and relative angle between coils $\theta$.  Note that this expression is not dependant on the size of the primary coil.  The following sections detail the calculation of coupling coefficients for 2D planar and solenoid coil geometries, where the coil height is much smaller, or much larger than the radius of the coil respectively.

\subsection{Planar Coil Approximation}
In this instance, $r'_{2a}$ and $r'_{2b}$ can be geometrically defined for any separation $a$ and relative angle $\theta$ (see Figure~\ref{fig:coil-geom}.c), as,
\begin{align}
     r'_{2a} &= a\sqrt{2 \left(1 -\cos \theta\right)},\label{eqn:r2a_gen}\\
     r'_{2b} &= r_2\sqrt{2\left[2 + 2\zeta\left(1 -\cos \theta\right) + \zeta^2 \left(1 -\cos \theta\right)\right]},\label{eqn:r2b_gen}
\end{align}
where $\zeta = {a}/{r_2}$ is defined as the \textit{separation ratio}. Note that for simplicity, here we have assumed that the normal component of the magnetic flux will be normal to the central plane of the coil thereby eliminating the need to resolve the components of the B-field. This simplifies the integration in equation~\ref{eqn:k_gen}.  The sections below discuss specific cases for how the coupling between coils will change as a function of different variables - co-planar separation ($a$) and angular displacement ($\phi = \pi - \theta$).

\hfill
\subsubsection{Co-planar Separation, $a$}\label{sec:coplanar}
When the coils are co-planar ($\theta=\pi$), as shown in Figure~\ref{fig:coil-geom}.a, Equations~\ref{eqn:r2a_gen}-\ref{eqn:r2b_gen} can be simplified to $r'_{2a}=2a$ and $r'_{2b}=2r_2(\zeta + 1)$. We can therefore define the coupling coefficient from equation~\ref{eqn:k_gen} as a function of the dimensionless separation ratio, $\zeta$,
\begin{equation}
k = \frac{1}{4\pi} \ln{\left( \frac{\zeta + 1}{\zeta} \right)}. \label{eqn:kl}
\end{equation}
The equation follows the expected $1/a$ form for the decay in magnetic field around a current carrying wire \cite{Griffiths2008IntroductionElectrodynamics}.
% For example, a realistic minimum pitch between the edges of two neighbouring coils is 0.1mm, which, for coils of radius $r_1 = r_2 = 0.5 mm$ is equivalent to $\zeta = 0.2$, and achieves a coupling coefficient of $k\approx0.19$. 
% For larger coils with the same minimum gap between coils (0.1 mm) as a manufacturing limitation the maximum coupling coefficient tends towards $k=0.5$, which is consistent with the symmetry of the system, such that only half of the magnetic field generated by coil 1 will pass through coil 2, while the other half will pass on the opposite side of coil 1 (see Figure~\ref{fig:coil-geom}).  
% However, this assumption is only valid when the coils are air-cored. As soon as 1 or both coils have magnetically permeable materials in their core, this assumption would be invalid as the magnetic field would be distorted towards regions of high magnetic permeability.
%angular displacement

\subsubsection{Angular Displacement, $\phi$}
When the two planar coils are no longer co-planar, i.e. $\theta < \pi$ ($\phi > 0$), the full expressions for $r_{2a}'$ and $r_{2b}'$ (equations~\ref{eqn:r2a_gen}-\ref{eqn:r2b_gen}) can be used to calculate the relationship between $\theta$ and the coupling coefficient.
In order to aid the inversion of the relative angle between the coils, a simplified first order expression for this relationship is derived in Appendix~\ref{App:C}. 
The coupling coefficient, $k$, can therefore be defined as,
\begin{align}
       k \approx \frac{1}{8\pi} \ln{\left[1 + \frac{2}{\zeta}+\frac{4}{\zeta^2\left(4 - \phi^2\right)}\right]},\label{eqn:k_ang_small}
\end{align}
where $\phi$ is given in radians. 
% An analytical expression for the displacement angle, $\phi$, between neighbouring filament coils can therefore be defined as,
% \begin{align}
%        \phi \approx 2\sqrt{1 - \frac{1}{2\zeta \left(e^{8\pi k} - 1\right)-1}}. \label{eqn:phi_ang_small}
% \end{align}
% These formulae predict that if the coupling coefficient, $k$, can be measured experimentally, then the displacement angle can be estimated. The final question to address is what is the predicted change in $k$ for non-planar coils i.e. finite height coils.

\begin{figure*}[!t]
\centering
\includegraphics[width=5.0in]{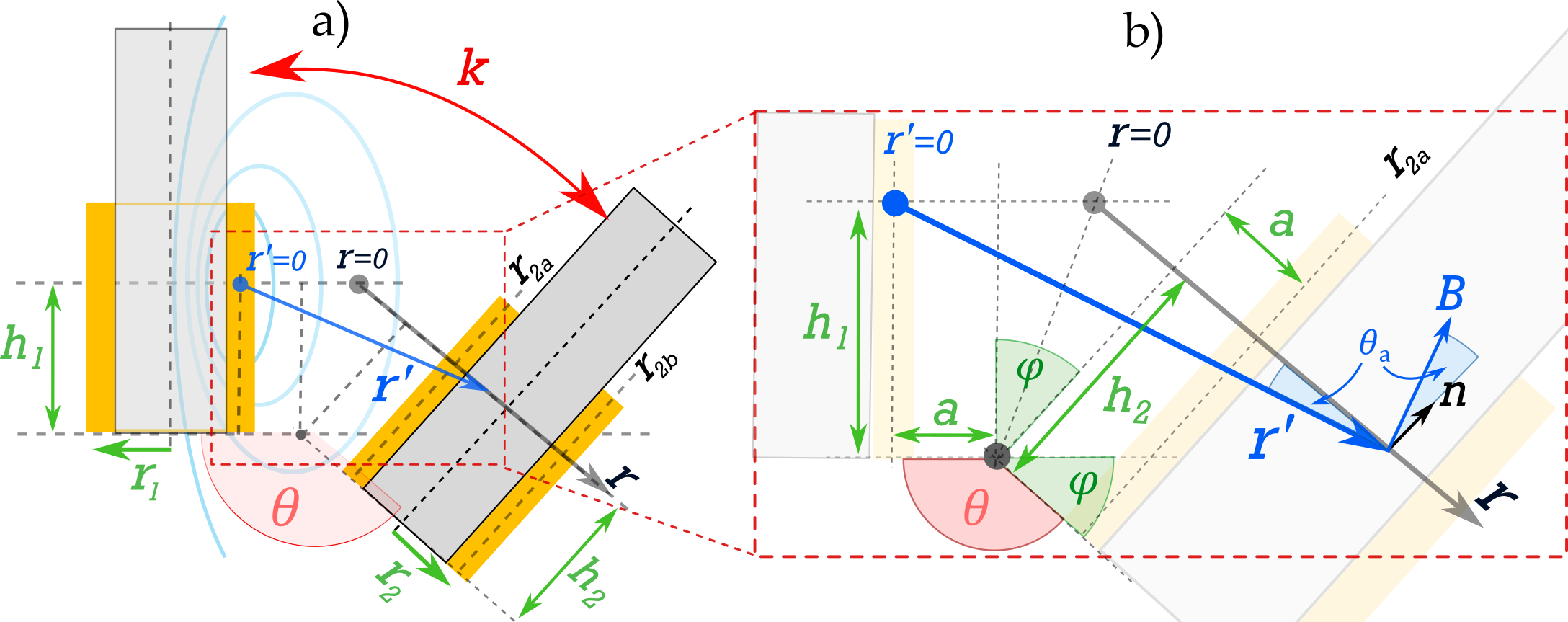}
\caption{Model configuration of semi-infinite ferrite cored solenoid coils, where the yellow blocks represent the cross section of the coil, showing; a) angular separation of the centre of magnetism of each coil, and b) zoomed in detail of trigonometric relationship between distance from the line of symmetry ($r=0$) and distance from centre of the B-field source (at $r'=0$) at a centre of magnetism height, $h$, within the coil windings.}
\label{fig:coil-geom-big}
\end{figure*}

\subsection{Solenoid Coil Approximation}
To approximate the behaviour of finite-height 2D coils, we make the assumption that the solenoid coils exhibit magnetic flux densities with north-south symmetry about a central plane (centre of magnetism - CoMag) at some height, $h$, along the coil axis.  Here we have assumed that the coils are identical such that $h_1=h_2=h$.  This is taken as the averaging plane of the sensor and is the plane along which the flux density will be integrated. If coils contain ferrite cores, these will act to shift the CoMag plane, depending on the relative height and location difference between coil windings, $h_w$, and core, $h_c$, (see Figure~\ref{fig:coil-geom-big}).

Unlike the planar coil equivalent in the previous section, increasing the angular rotation about a pivot point in the basal plane of the coils, increases the separation between the CoMag of the two coils, reducing the magnetic bridging between them. As such, the coupling coefficient between coils is expected to decrease with increasing angle, $\phi$. 
Recalculating $r_{2a}'$ and $r_{2b}'$ along the CoMag for coil 2 (assuming still that coils 1 and 2 are identical), we can arrive at general expressions,
\begin{align}
     r'_{2a} &= r_2 \Lambda \sqrt{2 \left(1 -\cos \theta\right)},\label{eqn:r2a_big}\\
     r'_{2b} &= r_2\sqrt{2\left[2 + 2\Lambda\left(1 -\cos \theta\right) + \Lambda^2 \left(1 -\cos \theta\right)\right]},\label{eqn:r2b_big}
\end{align}
where,
\begin{align}
\Lambda = \zeta + \eta\tan{\frac{\phi}{2}},\label{eqn:LAM}
\eta = \frac{h}{r_2}.
\end{align}
% Following a similar approach of simplification and first-order approximation  (see Appendix~\ref{App:D}), these can be approximated for the case when $a \ll h$, $\zeta \ll \eta$ , to,
% \begin{align}
%      r'_{2a} &\approx h\phi, \label{eqn:r2a_big_simp}\\
%      r'_{2b} &\approx r_2\left(2 + \eta \phi\right),\label{eqn:r2b_big_simp}
% \end{align}
% where the coil aspect ratio $\eta = \frac{h}{r_2}$.  

It is clear that, in the case when $\theta = \pi$ ($\phi = 0$), $r'_{2a}$ and $r'_{2b}$ simplify to the same expressions for the co-planar separation defined in section~\ref{sec:coplanar}, giving the same formula for $k$ as given in equation~\ref{eqn:kl}. Equations~\ref{eqn:r2a_big}-\ref{eqn:LAM} can therefore be considered the generalised formulae for calculating $k$ between neighbouring identical coils.

\hfill
\subsubsection{Angular Displacement, $\phi$}
Substituting these new expressions for $r'_{2a}$ and $r'_{2b}$ into equation~\ref{eqn:k_gen} allows us to calculate $k$ as a function of $\phi$ for increasing CoMag heights, $h$, as shown in Figure~\ref{fig:ksepRat}.b, from planar (blue) to solenoid (red) coils.  
For the case when $\zeta \ll \eta$ (i.e. $a \ll h$), $r_{2a}'$ and $r_{2b}'$ can be approximated as (see Appendix~\ref{App:D}),
\begin{align}
    r_{2a}' &\approx h\sqrt{2\left(1 - \cos{\phi} \right)},\\
    r_{2b}' &\approx r_2\sqrt{4 + 4\eta \sin{\phi} + 2\eta^2\left(1 - \cos{\phi}\right)}.
\end{align}
The first-order approximation of $k$ can therefore be defined as,
\begin{equation}
    k \approx \frac{1}{4\pi} \ln{\left[ 1 + \frac{2}{\eta\phi} \right]}. \label{eqn:k_big_approx}
\end{equation}
% The approximation predicts a $1/\phi$ (in radians) relationship which is again consistent with expectations for solenoid coils which 

\subsection{Predicted Trends}
\begin{figure*}[!h]
\centering
\includegraphics[width=6.0in]{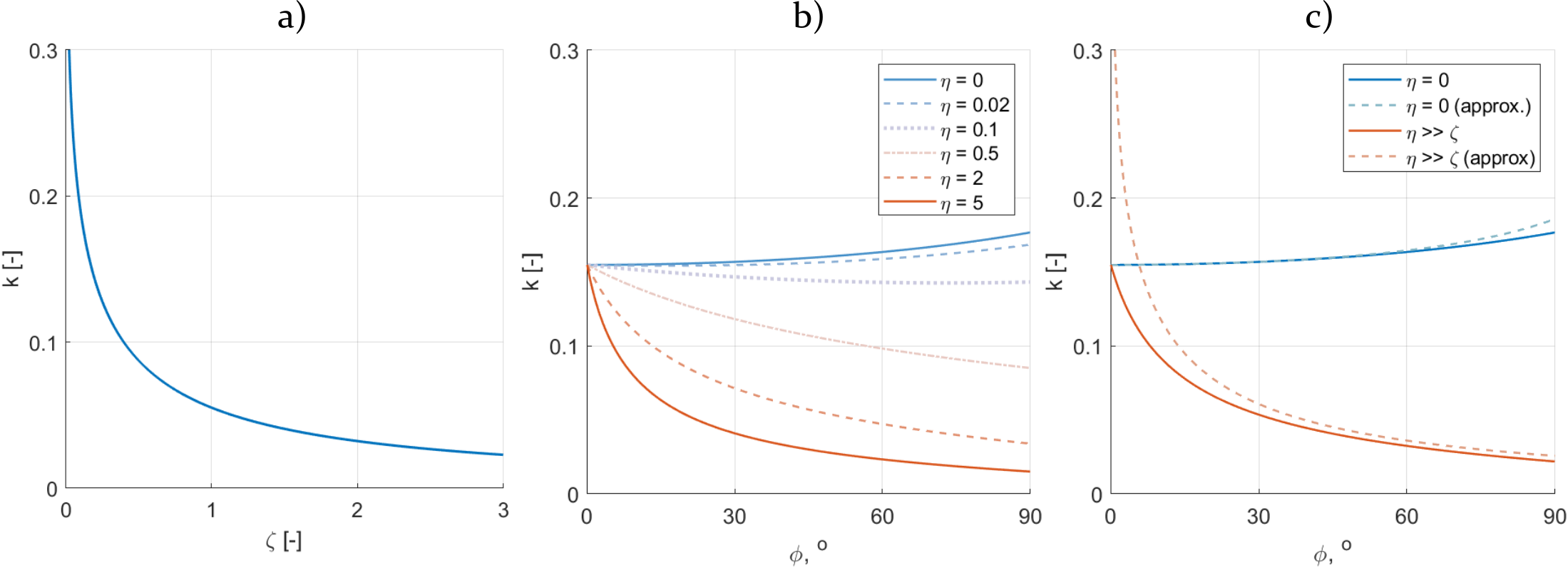}
% \includegraphics[width=2.0in]{df_k.png}&
% \includegraphics[width=2.15in]{df_zeta.png}\\
% a)&b)&c)\\
\caption{Simplified model for calculating coupling coefficients, $k$, for neighbouring 2D coils showing; a) $k$ as a function of separation ratio $\zeta$ in co-planar coils, b) shows how the trends in $k(\phi)$ change with the height-to-radius aspect ratio $\eta$, with changing height from planar (blue) to solenoid (red). c) Example $k$ as a function of angle from the plane, $\phi$, for planar (blue) and solenoid (red) type coils (from equations~\ref{eqn:kl} and~\ref{eqn:r2a_big}-\ref{eqn:LAM}) compared to the first-order approximations (equations~\ref{eqn:k_ang_small} and~\ref{eqn:k_big_approx}) of each scenario (dashed).}
\label{fig:ksepRat}
\end{figure*}

Figure~\ref{fig:ksepRat} shows the calculated coupling coefficients as a function of the displacement variables ($\zeta$ and $\phi$) for the planar and solenoid coils.  Figure~\ref{fig:ksepRat}.a shows how $k$ varies with the dimensionless separation ratio, $\zeta$, based on the approximation in equation~\ref{eqn:kl}.  This theoretical analysis demonstrates how rapidly the coupling decays as a function of separation between the two coils, and also indicates the anticipated coupling coefficient range for a planar coil pair.
% When the coils are as close as physically possible, $a$ is equal to the diameter of the wire used such that $\zeta \rightarrow 0$. 
% For the case when the coils are very close ($a \ll r_2$) the coupling between two neighbouring air-cored coils of semi-infinite length is calculated to be $k = 0.367$ \footnote{Interestingly the same value as $e^{-1}$.}.  
% We can rearrange equation~\ref{eqn:kl} to arrive at an expression for the separation ratio, $\zeta$, as a function of the coupling coefficient,
% \begin{equation}
% \zeta = \frac{a}{r_2} \approx \left(e^{4\pi k}-1\right)^{-1}, \label{eqn:k2}
% \end{equation}
% Therefore the separation ratio can be found from a measurement of the coupling coefficient between coils.
Figure~\ref{fig:ksepRat}.b shows $k$ calculated from Equations~\ref{eqn:k_gen}-\ref{eqn:r2b_gen} as a function of the angular displacement, $\phi = \pi - \theta$, for coils with increasing height $h_1=h_2=h$ (defined by their coil aspect ratio, $\eta$).  The results demonstrate that with only a relatively small height to radius ratio ($\eta>0.1$), the angular displacement changes from increasing steadily with $k$, to decreasing rapidly.  This prediction indicates that there exists a ratio, $\eta$, where $k$ changes minimally with angular displacement.  This may be a valuable design property for many applications.

Figure~\ref{fig:ksepRat}.c compares the full calculation from equations~\ref{eqn:k_gen}-\ref{eqn:r2b_gen} (solid lines) to the trends predicted by the first order approximate formulae (dashed lines) derived for the 2 extreme cases, planar (blue lines) and solenoid (red lines) coils. 
% The blue curves show the difference between this first-order approximation and the full formula expression.  The approximation shows excellent agreement with the full calculation for values of $0<\phi<\pi/3$.  The predicted $k$ is shown in Figure~\ref{fig:ksepRat}.c as a function of $\phi$ for the full simplified model calculation in comparison to the first-order approximation formula of equation~\ref{eqn:k_big_approx} (red solid and dashed lines respectively).  
The first order approximation shows the same trend as the full formula, but diverges at low values of $\phi$, due to the assumption $a\ll h$, while showing excellent agreement for values of $\phi > 30^\circ$.  
% As before,  an expression can be defined for $\phi$ for a measured value of $k$ as,
% \begin{equation}
%     \phi \approx \frac{2}{\eta e^{4\pi k} - \eta}.
% \end{equation}
% The predicted behaviours demonstrated in Figure~\ref{fig:ksepRat} were evaluated using 2D and 3D finite element simulations for the two scenarios; 2D planar-type coil (PCB), and solenoid coils.

% Figure~\ref{fig:ksepRat} were evaluated using 2D and 3D finite element simulations for the two scenarios; 2D filament-type coil (PCB), and 2D solenoid-type coils.

\section{Finite Element Modelling}
The approximate formulae derived above are only valid for the condition when a coil can be approximated to a 2D coil. Finite element (FE) simulations of coils in 2D are unable to predict parameters such as inductance, capacitance or resistance of the coil, however they can be used to quantitatively predict the magnetic flux density surrounding neighbouring coils in order to determine the expected flux sharing (i.e. coupling coefficient) between elongated coils. The magnetic flux in the region around a 2D simulation can be presumed valid if a coil is sufficiently elongated in the out-of-plane (z-axis) direction \cite{Hrabovsky2019TheMultiphysics}. Magneto-static coil models were developed in 2D and 3D with the AC/DC module in COMSOL Multiphysics 6.1 (see Figure~\ref{optimisation}.a-b) and used to evaluate the impact of coil geometry on the magnetic flux around a driver coil. The models simulate a single winding layer solenoid coil with winding height $h_w$, core height from the basal plane $h_c$, core diameter $d_c$, coil length $l_c$ (along the z-axis for 3D models), and a core relative magnetic permeability of $\mu_r = 25$ to represent typical values for iron ferrite cores \cite{Kaverine2017}.

\subsection{2D Model Approximation}
\begin{figure*}[!b]
    \centering
    \includegraphics[width=\textwidth]{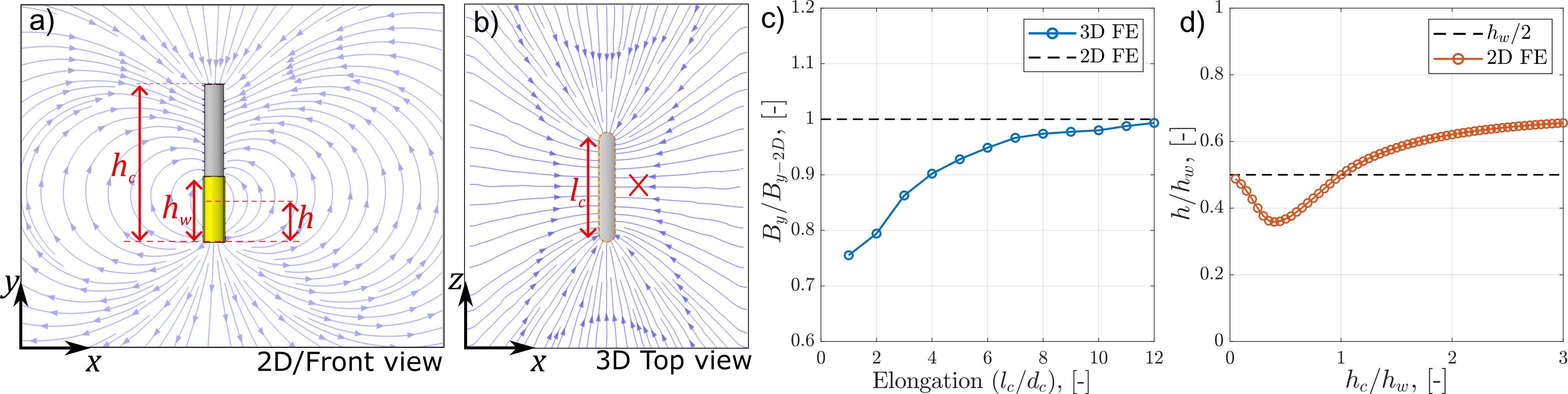}
    \caption{2D FE model convergence: a) 2D model / 3D cross-sectional view of a simulated coil of core height, $h_c$, winding height, $h_w$, and plane of symmetry of the magnetic field (centre of magnetism), $h$. b) 3D top down view showing coil length, $l_c$. The red X indicates the evaluation point of the magnetic flux density ($B_y$) as a function of coil elongation (ratio between the coil length, $l_c$, and diameter, $d_c$) with the results shown in c) relative to the flux predicted by the equivalent 2D FE model ($B_{y-2D}$). Graph d) plots the centre of magnetism $h$ relative to the coil winding height $h_w$ as a function of the core and winding height.}
    \label{optimisation}
\end{figure*}

In order to validate and compare to the coupling coefficients predicted in section~\ref{sec:kmodel}, a virtual study was conducted to determine at what aspect ratio of coil length, $l_c$, to diameter, $d_c$, a 3D coil will begin to behave like an infinite 2D coil.  A 3D FE model was employed to evaluate the in-plane vertical component of the magnetic flux density, $B_y$, at a point next to an excitation coil, as a function of $l_c$.  The relative convergence of the 3D and 2D model values for $B_y$ is shown in dimensionless form in figure~\ref{optimisation}.c. Models were generated with nominal coil dimensions of $d_{c}=10~mm$, $h_w=20~mm$ and $h_c=50~mm$.  

The results show that for an elongation ratio of 5, the 3D model predicts $B_y$ at the centre plane of the coil to be $93\%$ of $B_{y-2D}$. At an elongation ratio of 10, the 3D models B-field reaches $98\%$ of $B_{y-2D}$. While the B-field along the central plane of the coil in 3D is comparable to the 2D model, the B-field along the full length of the coil will not be the same as the 2D model. However, this virtual study indicates the minimum coil aspect ratio to begin physically approximating a 2D coil is $l_c > 5d_c$.

\subsection{Centre of Magnetism}
The simple analytical models developed in section~\ref{sec:kmodel} are based upon an assumption that the magnetic field from a coil can be modelled as circular, being emitted from a magnetic point source within the coil. This point source is assumed to be centred at the mean of the winding radius, and at a height $h$ from the basal plane. For a coil perfectly coaxial with a core equal in height to the coil windings ($h_c=h_w$), or if the coil is air-cored, then the centre of magnetism (CoMag) would be expected to be half way up the coil windings ($h_w/2$), as shown in Figure~\ref{optimisation}.d.  This is confirmed by finding the plane of the turning point of the B-field along the $x$-axis i.e. the point at which the $x$-component of the B-field $B_x$ passes through zero.  

The results shown in figure~\ref{optimisation}.d demonstrate that as the core increases in height inside the coil windings, the CoMag moves below $h_w/2$ as it is pulled towards the high permeability core at the base of the windings.  The CoMag ($h$) goes through a minimum at $h_c = 0.4h_w$ where $h\approx0.36h_w$.  The CoMag then passes through $h_w/2$ when $h_c = h_w$ before tending to value of $h \approx 0.67h_w$, with coil cores $h_c > 2h_w$ no significant changes in $h$ is observed.

The following section compares the coupling coefficients predicted by the 2D FE models to the first order approximations developed in section~\ref{sec:kmodel}, as a function of co-planar separation and relative angle.

\section{Validation of Approximation}
The magneto-static 2D FE models were extended to simulate the magnetic flux sharing between two identical coils (one driver, one passive) during separation and angular displacement. The coupling coefficient $k$ between the coils was evaluated by analysing the flux through the CoMag of each coil and taking the ratio as defined by equation~\ref{eqn:k_flux0}.  Only coils with equal core and coil winding heights were considered such that the CoMag was in the mid-plane of each coil (see Figure~\ref{optimisation}.d).

\begin{figure*}[!t]
    \centering
    \includegraphics[width=6.0in]{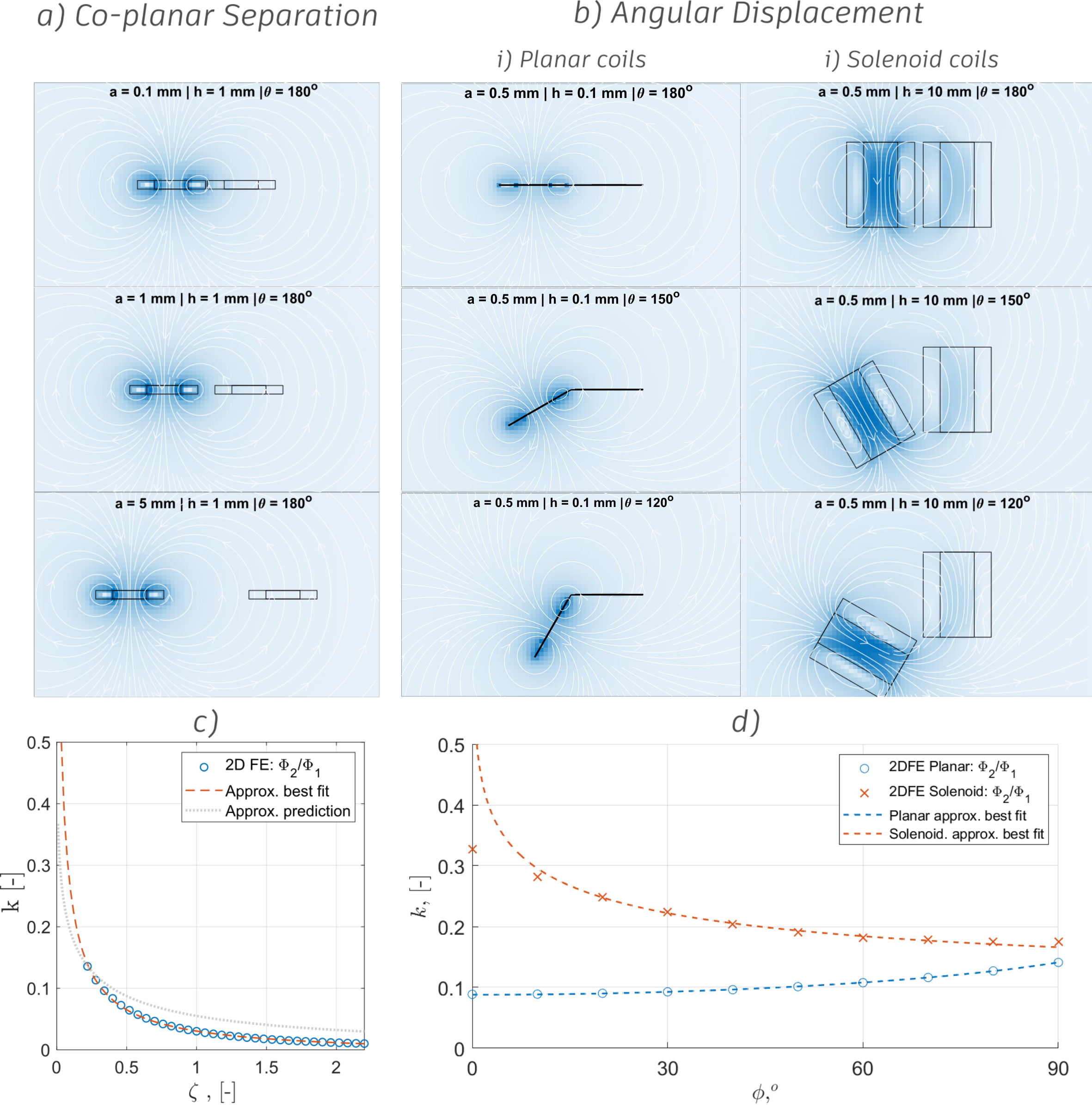}
    \caption{2D FE predictions - Simulating the static magnetic flux generated by a driver coil, and the amount of flux experienced by an identical neighbouring coil, as a function of a) co-planar separation ($a$), and b) angular displacement $\phi$. Graphs c) and d) show the the 2D FE predicted coupling coefficients, $k$, as a function of separation ratio ($\zeta = a/r_2$), and displacement angle, $\phi$, compared to the best fit curves generated via the 1st order approximations. c) compares to the directly calculated function from equation~\ref{eqn:kl} (grey dashed line).}
    \label{fig:FEresults}
\end{figure*}

Figure~\ref{fig:FEresults} shows the results of the virtual studies calculating $k$ from simulated magneto-static flux densities as a function of co-planar separation (Figure~\ref{fig:FEresults}.a and c) and angular displacement (Figure~\ref{fig:FEresults}.b and d). Both planar and solenoid coils (Figure~\ref{fig:FEresults}.b.i and ii respectively) are evaluated as a function of displacement angle, $\phi$ to highlight differences in trends between the 2 extreme coil geometries, while only a planar coil is considered for co-planar separation, as the trends with respect to the separation ratio are consistent for both co-planar and solenoid coils.  The coils each had a core radius $2~mm$, and winding thickness of $2~mm$, with the a turn density of $100~mm^{-1}$. The angular displacement simulations were conducted with each coil a distance $a = 0.5~mm$ from the pivot point on the basal plane of the coils.

The results shown in Figure~\ref{fig:FEresults}.c and d demonstrate that the first order approximations derived in section~\ref{sec:kmodel} accurately predict the trends in coupling coefficient between elongated coils as verified by 2D FE modelling between.  

Figure~\ref{fig:FEresults}.c shows the calculated coupling coefficient as a function of separation ratio, $\zeta$ (grey dashed line) from equation~\ref{eqn:kl}, which predicts comparable trends to the FE model. The model is therefore inaccurate in predicting explicit values for $k$.  This is to be expected given the number of assumptions made about the system during the derivation of the formula for $k$. However, the formula given in equation~\ref{eqn:kl} can be used to plot a best fit function against FE data.  

This is done by rearranging equation~\ref{eqn:kl} to give a linear expression and plotting the terms against one another, in this case, if $x=1/\zeta$ and $y=e^{\pi k}$, then $y=Ax+B$ and a linear fit can be found to determine the unknown gradient and intercept ($\alpha$ and $\beta$ respectively); see Table~\ref{table:bestFit}. It should be noted that the factor of 4 from equation~\ref{eqn:kl} is omitted as it was found to result in a stronger linear correlation.  This is most likely due to the assumptions around the circular nature of the emitted field.  However, the resulting best fit curve provides an extremely high correlation with FE results (red dashed line in Figure~\ref{fig:FEresults}.c).

The same process of first-order approximation curve fitting was applied to both the planar and solenoid coil FE results for angular displacement with the results in Figure~\ref{fig:FEresults}.d showing the strong correlation achieved with this simple approximation between $0^o<\phi<90^o$.  The only deviation occurs at low $\phi$ for the solenoid coil as the function tends to infinity.  Both angular displacement curves retained the factors of 8 \& 4 from equations~\ref{eqn:k_ang_small} and \ref{eqn:k_big_approx} respectively to achieve the optimum fit (see Table~\ref{table:bestFit}).

Table~\ref{table:bestFit} summarises the first-order approximation expressions used to fit to the 2D FE model results, and provides the best-fit coefficients for the fitted curves shown in Figure~\ref{fig:FEresults}.c \& d.  The resulting fitted curves exhibit standard deviations from the 2D FE data of $\pm 2\%$ for co-planar separation, and $\pm 0.4\%$ and $\pm 2.1\%$ for angular displacement of the planar and solenoid coils respectively.

\begin{table*}[ht!]
\centering 
\caption{1st order approximation best fit expressions and fitted coefficients compared to 1st order model predicted coefficients}
\label{table:bestFit}
\begin{tabular}{ l l c c } 
\hline
  & {Best fit expression} & \textbf{$\alpha$} & \textbf{$\beta$} \\
\hline
 $\zeta$ (Eqn.~\ref{eqn:kl}) & $ k = \frac{1}{\pi}\ln{\left[\alpha\frac{1}{\zeta}+\beta\right]}$ & 0.13 & 0.97 \\  
$\phi$: Plan. (Eqn.~\ref{eqn:k_ang_small}) & $k = \frac{1}{8\pi}\ln{\left[\alpha\left(\frac{1}{4-\phi^2}\right)+\beta\right]}$ & 63.56 & -6.76 \\
$\phi$: Sol. (Eqn.~\ref{eqn:k_big_approx})& $k = \frac{1}{4\pi}\ln{\left[\alpha\left(\frac{1}{\phi}\right)+\beta\right]}$ & 5.20 & 5.61 \\
  \hline
\end{tabular}
\end{table*}

\subsection{Model Limitations}
The simplified 2D coupling coefficient models developed in this paper demonstrate excellent agreement with the trends produced using 2D FE models for both co-planar separation, and for angular displacement within the range $\phi = 0-90^o$, but is limited to this angular range.

Due to the approximations made, these simple formulae are unable to accurately predict explicit values for the coupling coefficients given specific input parameters.  This is demonstrated in table~\ref{table:vars} where the fitted coefficients for the angular displacement curves are used to invert the physical variables of the system, separation and aspect ratios ($\zeta$ and $\eta$ respectively) from the equations in section~\ref{sec:kmodel}.

\begin{table*}[t!]
\centering 
\caption{Calculated variables from first order fitted coefficients, showing predicted values from geometric formulae in brackets.}
\label{table:vars}
\begin{tabular}{ l c c c c} 
%  & & & \multicolumn{2}{|c|}{calculated vars. (geometrically predicted)} \\
% \hline
   & \textbf{$\alpha$} & \textbf{$\beta$} & &  \\
\hline
$\phi$: Plan. ($\zeta=a/r_2$)& $4/\zeta^2 $ & $1 + (2/\zeta)$ & $\zeta \approx \pm 0.25\ (\textit{0.17})$ & $\zeta \approx -0.26\ (\textit{0.17})$\\
$\phi$: Sol. ($\eta=h/r_2$) & $2/\eta $ & $-$ & $\eta \approx 0.38\ (\textit{1.67})$ & $ - $\\
  \hline
\end{tabular}
\end{table*}

The table compares the calculated values to the predicted ratios, as determined from the geometric properties of the coil system, assuming that the centre of the magnetic field is in the middle of the coil windings as $r_2 = 3~mm$.  The results show the deviation between the simplified model prediction and the calculated values demonstrating the limitations in the prediction capabilities of this first order approach.  This is not surprising giving the many assumptions that were employed to reach the formulae derived. However, future work may find insight by evaluating the separation and coil aspect ratios in order to inform the geometric location $(a,h)$ of the equivalent magnetic point sources for the planar and solenoid first order models. 

There are many different factors that have been omitted from the model for simplicity in order to arrive at simple formulae for easy fitting which have been shown to be sufficient for the extreme coil geometry cases proposed.  Factors include variation in magnetic permeability with different core aspect ratio \cite{Kaverine2017}, and the likely distortion of the CoMag as a function of angular displacement.

\subsection{Comparison to Existing 3D Models}
\begin{figure*}[!h]
    \centering
    \includegraphics[width=4.0in]{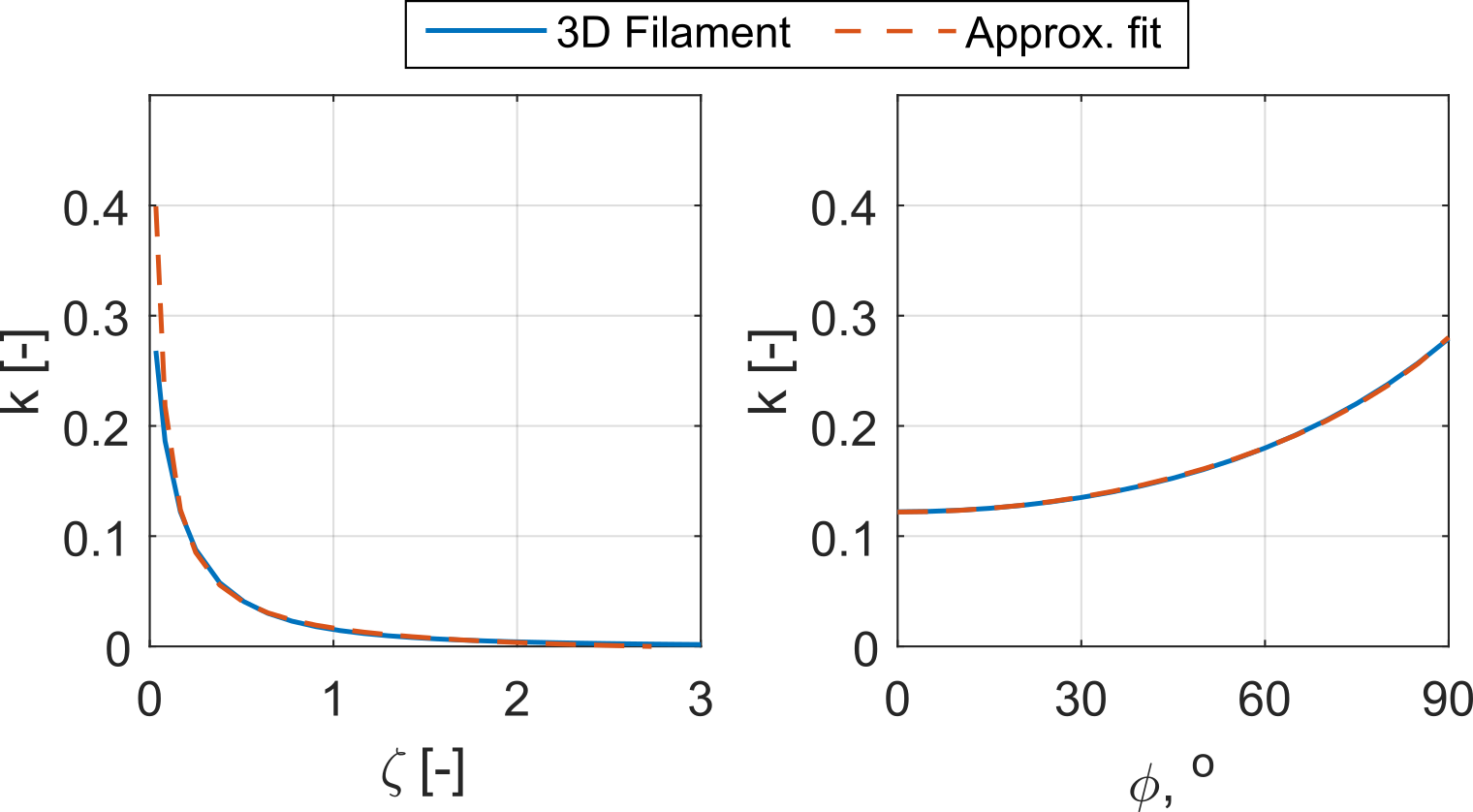}
    \caption{3D coil predictions - Showing plots of the coupling coefficient, $k$, as a function of; a) coplanar separation ratio, $\zeta = a/r_2$, and b) angular displacement, $\phi$, in degrees. Showing $k$ calculated numerically using a 3D filament model (blue solid lines) from \cite{Poletkin2019EfficientMethod} for coils of radius $r_n = 3~mm$ and separation of $a = 0.5~mm$, compared to the best-fit (red dashed lines) using the first-order approximation function derived herein.}
    \label{fig:3dpred}
\end{figure*}

The proposed first-order model exhibits comparable trends in coupling coefficient predictions as those calculated using the numerically solved formulae provided in the supplementary data by Poletkin \& Korvink \cite{Poletkin2019EfficientMethod} for 3D circular filaments coils.  While the exact conversion from mutual inductance to coupling coefficient requires the determination of the self-inductance of the coils, the predicted coupling coefficients (for a nominal coil inductance) follow comparable trends with separation ratio and increasing angle given by this first-order model proposed here (shown in Figures~\ref{fig:3dpred}). While the absolute values and rates of decay will not match, this mismatch is due to the difference between the 2D approximation verses the 3D numerical model.  However, the first-order approximation formulae remain applicable for fitting to coupling coefficients calculated using 3D coil numerically solved formulae from \cite{Poletkin2019EfficientMethod}, as shown in Figure~\ref{fig:3dpred}.

The results shown in Figure~\ref{fig:3dpred} demonstrate the versatility of the approximate model for function fitting to 3D scenarios.  
Table~\ref{table:bestFit_3d} gives the approximate functions and coefficients plotted in Figure~\ref{fig:3dpred} for filament coils of radius $3~mm$ and separation from the pivot point of $a=~0.5mm$. These values (in comparison to those for the 2D case in Table~\ref{table:bestFit}) highlight the differences in divergence of the magnetic field around a 3D coil compared to the 2D approximation. 
It is noted that the best fit for the angular displacement of the 3D filament case is found when the term $1/8\pi$ (from equation~\ref{eqn:k_ang_small}) becomes $2/5\pi$. 
This change is based on empirical observation, and not a physics definition, and so little physical meaning should be assigned to this change, other than that there is a difference between the 2D and 3D cases. 
The coefficients $\alpha$ and $\beta$, for the 3D angular displacement case, are found to be around $1/4$ of their respective coefficients for the 2D planar coil case (Table~\ref{table:bestFit}). 
The coplanar separation case shows that coefficient $\beta$ is almost unchanged between the 2D and 3D cases, while $\alpha$ is approximately $70\%$ of $\alpha$ for the 2D case.
This result demonstrates that these physics-based first-order functions can be employed as effective approximations for fitting to 3D problems to enable simple inversion of physical separation between $0-90^o$. 
% The resulting fitted function exhibit a standard deviation of $<0.5\%$.

% 26.7811   -3.1519
%     4.9029    0.7123

\begin{table*}[!h]
\centering 
\caption{First-order approximation best fit expressions and fitted coefficients for fitting of 3D filament model in \cite{Poletkin2019EfficientMethod}.}
\label{table:bestFit_3d}
\begin{tabular}{ l l c c } 
\hline
  & {Best fit expression} & \textbf{$\alpha$} & \textbf{$\beta$} \\
\hline
%  $\zeta$ (Eqn.~\ref{eqn:kl}) & $ k = \frac{1}{\pi}\ln{\left[\alpha\frac{1}{\zeta}+\beta\right]}$ & 0.024 & 0.99 \\  
% $\phi$: Plan. (Eqn.~\ref{eqn:k_ang_small}) & $k = \frac{1}{8\pi}\ln{\left[\alpha\left(\frac{1}{4-\phi^2}\right)+\beta\right]}$ & 166.59 & -28.13 \\
 $\zeta$ (Eqn.~\ref{eqn:kl}) & $ k = \frac{1}{\pi}\ln{\left[\alpha\frac{1}{\zeta}+\beta\right]}$ & 0.08 & 0.97 \\ 
$\phi$: Plan. (Eqn.~\ref{eqn:k_ang_small}) & $k = \frac{2}{5\pi}\ln{\left[\alpha\left(\frac{1}{4-\phi^2}\right)+\beta\right]}$ & 15.99 & -1.39 \\
  \hline
\end{tabular}
\end{table*}

\section{Conclusions}
A novel set of equations have been derived using simple geometric arguments and first-order approximations to predict the trends expected in the coupling coefficients between neighbouring identical coils as a function of relative angle and co-planar displacement.
 While the equations proposed are not suitable for accurate forward calculation of coupling coefficients, the formulae accurately predict the trends simulated in magneto-static 2D FE models, and can be used to fit to coupling coefficient measurements with exceptional agreement between $0-90^o$ to both 2D FE and 3D numerical models, thereby enabling the possibility of direct physical inversion of the relative displacement between coils, based on a fitted calibration curve.
It is expected that these 2D formulae have applications in the fields of wireless power transfer, electro-mechanical motor design, and inductive sensing measurement, as a fast and accurate technique for enabling experimental calibration of flux sharing multi-coil sensing systems. Moreover the formulae could be employed for simple coil geometry optimisation either to enhance or minimise the effects of relative changes in angle between neighbouring coils.
Future work will experimentally confirm the validity of these expressions and apply them for the direct inversion of displacement between coils.

% \begin{figure}[!t]
%     \centering
%     \includegraphics[width=3.0in]{Images/k_theta_a500_Hc_mu25.png}
%     \caption{FE model results for increasing coil height}
%     \label{optimisation}
% \end{figure}

% \begin{figure}[!ht]
%     \centering
%     \includegraphics[width=4.0in]{Images/Kpcbcomparison.eps}
%     \caption{Optimisation of the coil; a)Front view of the coil where the height is increased and evaluated in three different points where the position of the receiver coil, b) the contribution of the magnetic flux where the black line is the objective variable, c) magnetic flux field from the top view, d) increment of the magnetic flux in the $Y$ component with an increasing length}
%     \label{optimisation}
% \end{figure}

\appendix

\section{Angular Displacement Approximation}\label{App:C}
From the general expressions for the radial distance to the nearest and furthest edges of coil 2 (equations~\ref{eqn:r2a_gen}-\ref{eqn:r2b_gen}), and recognising that $\cos{\theta}=\cos{(\pi - \phi)} = - \cos{\phi}$, we can define,
\begin{align}
     r'_{2a} &= a\sqrt{2 \left(1 +\cos{\phi}\right)},\label{eqn:r2a_gen_app}\\
     r'_{2b} &= r_2\sqrt{2\left[2 + 2\zeta\left(1 +\cos{\phi}\right) + \zeta^2 \left(1 +\cos{\phi} \right)\right]}.\label{eqn:r2b_gen_app}
\end{align}
Using the Taylor series of the cosine term up to $\mathcal{O}(\phi^2)$ gives,
\begin{align}
     r'_{2a} &\approx a\sqrt{4-\phi^2},\label{eqn:r2a_app_ang}\\
     r'_{2b} &\approx r_2\sqrt{4 + 2\zeta\left(4-\phi^2 \right) + \zeta^2\left(4-\phi^2 \right)}.\label{eqn:r2b_app_ang}
\end{align}
%$r'_{2b}$ can be factored to give,
% \begin{align}
%      r'_{2b} &\approx r_2\sqrt{4 + 4\left(1 + \zeta \right)^2 - \zeta\phi^2\left(2 + \zeta\right)},
% \end{align}
% which, when $a \ll r_2$, $\zeta \ll 1$, simplifies to,
% \begin{align}
%      r'_{2b} &\approx r_2\sqrt{8 - 2\zeta\phi^2}.\label{eqn:r2b_app_ang}
% \end{align}
Substituting equations~\ref{eqn:r2a_app_ang} and ~\ref{eqn:r2b_app_ang} into equation~\ref{eqn:k_gen} gives,
\begin{align}
    k &\approx \frac{1}{4\pi} \ln{\left[ \sqrt{\frac{r_2^2}{a^2}\frac{\left[4 + 2\zeta\left(4-\phi^2 \right) + \zeta^2\left(4-\phi^2 \right)\right]}{\left( 4-\phi^2 \right)} } \right]}, \\
    &= \frac{1}{8\pi} \ln{\left[ 1 + \frac{2}{\zeta} + \frac{4}{\zeta^2\left(4-\phi^2\right)}\right]}.\label{eqn:k_ang_1}%\\
    %&\approx \frac{1}{8\pi} \ln{\left[ \frac{2}{\zeta} + \frac{4}{\zeta^2\theta^2} \right]}. 
\end{align}

\section{Solenoid Centre of Magnetism Calculations}\label{App:D}
From Figure~\ref{fig:coil-geom-big}, we can define the radial distance from the magnetic field source, $r'=0$, at the nearest and further side of the windings for the second coil in 2D as,
\begin{align}
     r'_{2a} &= \left(a+h\tan{\frac{\phi}{2}}\right)\sqrt{2 \left(1 -\cos \theta\right)},\label{eqn:r2a_big_app}\\
     r'_{2b} &= r_2\sqrt{2\left[2 + 2\left(\frac{a+h\tan{\frac{\phi}{2}}}{r_2}\right)\left(1 -\cos \theta\right) + \left(\frac{a+h\tan{\frac{\phi}{2}}}{r_2}\right)^2 \left(1 -\cos \theta\right)\right]},\label{eqn:r2b_big_app}
\end{align}
where $h=h_1=h_2$ is the centre of magnetism plane in coils 1 and 2 and $\phi$ is the angular displacement from co-planar (i.e. $\phi=\pi-\theta$).  Via the trigonometric identities, $\tan{\frac{\phi}{2}} = (1-\cos{\phi})/\sin{\phi}$,  $\cos{\phi}=\cos{(\pi - \theta)} = - \cos{\theta}$, and $\sin{\phi} = \sin{(\pi - \theta)} = \sin{\theta}$ we have,
\begin{align}
    r_{2a}' &= \left[a + h\frac{\left(1 - \cos{\phi} \right)}{\sin{\phi}}\right]\sqrt{2\left(1 + \cos{\phi}\right)},\\
    &= a\sqrt{2\left(1 + \cos{\phi}\right)} + h\sqrt{2\frac{\left(1 - \cos{\phi} \right)^2\left(1 + \cos{\phi}\right)}{\sin^2{\phi}}},\\
    &= a\sqrt{2\left(1 + \cos{\phi}\right)} + h\sqrt{2\left(1 - \cos{\phi} \right)}.
\end{align}
Applying the first-order cosine Taylor series approximation used in section~\ref{App:C}, we can reach the approximation,
\begin{align}
    r_{2a}' &\approx a\sqrt{4 - \phi^2} + h\phi.
\end{align}
Setting $\eta=h/r_2$ and using the same trigonometric identities as before, an expression for $r_{2b}'$ can be given by,
\begin{align}
     r'_{2b} &= r_2\sqrt{\left[ 4 + 4\zeta\left(1 +\cos \phi\right) + 4\eta\sin{\phi} + 2\zeta^2\left(1 + \cos \phi\right) + 2\eta^2\left(1 - \cos \phi\right) + 4\zeta\eta\sin{\phi}\right]}.\label{eqn:r2b_big_a}
\end{align}
Applying the first-order sine and cosine Taylor approximation, we can simplify $r_{2b}'$ to,
\begin{align}
     r'_{2b} &\approx r_2\sqrt{\left[ 4 + 2\zeta(4-\phi^2) + 4\eta\phi + \zeta^2(4-\phi^2) + \eta^2\phi^2 + 4\zeta\eta\phi\right]},\\
     &= r_2\sqrt{\left[ 4(\zeta + 1)^2 + \phi^2(\eta^2 - \zeta^2 - 2\zeta) + 4\eta\phi(1 + \zeta) \right]}.\label{eqn:r2b_big_b}
\end{align}
Finally, in the limiting case when $\eta\gg\zeta$ (i.e. $h\gg a$), and $\zeta \ll 1$, then,
\begin{align}
     r_{2a}' &\approx h\phi,\\
     r_{2b}' &\approx r_2\sqrt{\left[ 4 + 4\eta\phi + \eta^2\phi^2 \right]},\\
     &= r_2 \left(2 + \eta\phi\right). \label{eqn:r2b_big_b}
\end{align}
Using equation~\ref{eqn:k_gen}, $k$ can be derived as,
\begin{align}
    k &\approx \frac{1}{4\pi} \ln{\left[ \frac{r_2\left[ 2 + \eta\phi\right]}{h\phi}  \right]} = \frac{1}{4\pi} \ln{\left[ \frac{ 2 + \eta\phi}{\eta\phi} \right]}=\frac{1}{4\pi} \ln{\left[ 1 + \frac{2}{\eta\phi} \right]}. \label{eqn:k_ang_1}
\end{align}

\section{Acknowledgements}
Alexis Hernandez's research is funded by the Consejo Nacional de Ciencia y Tecnología (CONACYT).
%\bibliographystyle{alpha}

% \bibliography{references2.bib}
\printbibliography

\end{document}